# Track Layouts of Graphs [*]

Vida Dujmović [†][‡]   Attila Pór [§]   David R. Wood [‡][§]

November 4, 2003
Revised August 15, 2018

**Abstract**

A $(k,t)$-*track layout* of a graph $G$ consists of a (proper) vertex $t$-colouring of $G$, a total order of each vertex colour class, and a (non-proper) edge $k$-colouring such that between each pair of colour classes no two monochromatic edges cross. This structure has recently arisen in the study of three-dimensional graph drawings. This paper presents the beginnings of a theory of track layouts. First we determine the maximum number of edges in a $(k,t)$-track layout, and show how to colour the edges given fixed linear orderings of the vertex colour classes. We then describe methods for the manipulation of track layouts. For example, we show how to decrease the number of edge colours in a track layout at the expense of increasing the number of tracks, and vice versa. We then study the relationship between track layouts and other models of graph layout, namely stack and queue layouts, and geometric thickness. One of our principle results is that the queue-number and track-number of a graph are tied, in the sense that one is bounded by a function of the other. As corollaries we prove that acyclic chromatic number is bounded by both queue-number and stack-number. Finally we consider track layouts of planar graphs. While it is an open problem whether planar graphs have bounded track-number, we prove bounds on the track-number of outerplanar graphs, and give the best known lower bound on the track-number of planar graphs.

---

[*]A preliminary version of this paper appeared as Technical Report TR-2003-07, School of Computer Science, Carleton University, Ottawa, Canada.

[†]School of Computer Science, McGill University, Montréal, Canada.

[‡]School of Computer Science, Carleton University, Ottawa, Canada. Research supported by NSERC.

[§]Department of Applied Mathematics, Charles University, Prague, Czech Republic. Research supported by COMBSTRU.

[¶]Email: vida@cs.mcgill.ca, por@kam.mff.cuni.cz, davidw@scs.carleton.ca

# Contents

# 1  Introduction

In its simplest form, a *track layout* of a graph consists of a vertex colouring and a total order on each colour class, such that there is no pair of crossing edges between any two colour classes. The purpose of this paper is to develop the beginnings of a theory of track layouts. Our focus is on methods for the manipulation of track layouts, and the relationship between track layouts and other models of graph layout. We consider undirected, finite, and simple graphs $G$ with vertex set $V(G)$ and edge set $E(G)$. The number of vertices and edges of $G$ are respectively denoted by $n = |V(G)|$ and $m = |E(G)|$.

A *vertex $|I|$-colouring* of a graph $G$ is a partition $\{V_i : i \in I\}$ of $V(G)$ such that for every edge $vw \in E(G)$, if $v \in V_i$ and $w \in V_j$ then $i \neq j$. The elements of $I$ are *colours*, and each set $V_i$ is a *colour class*. Suppose that $<_i$ is a total order on each colour class $V_i$. Then each pair $(V_i, <_i)$ is a *track*, and $\{(V_i, <_i) : i \in I\}$ is an $|I|$-*track assignment* of $G$. To ease the notation we denote track assignments by $\{V_i : i \in I\}$ when the ordering on each colour class is implicit.

An *X-crossing* in a track assignment consists of two edges $vw$ and $xy$ such that $v <_i x$ and $y <_j w$, for distinct colours $i$ and $j$. An *edge $k$-colouring* of $G$ is simply a partition $\{E_i : 1 \leq i \leq k\}$ of $E(G)$. A $(k,t)$-*track layout* of $G$ consists of a $t$-track assignment of $G$ and an edge $k$-colouring of $G$ with no monochromatic X-crossing; that is, edges of the same colour do not form an X-crossing. A graph admitting a $(k,t)$-track layout is called a $(k,t)$-*track graph*. The minimum $t$ such that a graph $G$ is a $(k,t)$-track graph is denoted by $\mathsf{tn}_k(G)$.

$(1,t)$-track layouts (that is, with no X-crossing) are of particular interest due to applications in three-dimensional graph drawing (see below). A $(1,t)$-track layout is called a *t-track layout*. A graph admitting a $t$-track layout is called a *t-track graph*. The *track-number* of $G$ is $\mathsf{tn}_1(G)$, simply denoted by $\mathsf{tn}(G)$. Dujmović *et al.* [25, 26] first introduced track layouts, although similar structures are implicit in many previous works [33, 40, 41, 51].

The graphs that admit 2-track layouts are easily characterised as follows, where a *caterpillar* is a tree such that deleting the leaves gives a path, as illustrated in Figure 1.

**Lemma 1.** [38] *A graph has a 2-track layout if and only if it is a forest of caterpillars.*

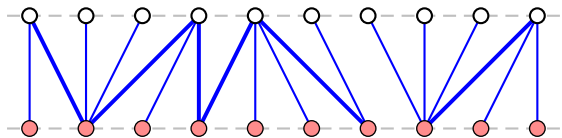

Figure 1: A 2-track layout of a forest of caterpillars.

Table 1 summarises the known bounds on the track-number.

Part of the motivation for studying track layouts is a connection with three-dimensional graph drawings. A *three-dimensional straight-line grid drawing* of a graph, henceforth called a *3D drawing*, is a placement of the vertices at distinct points in $\mathbb{Z}^3$ (called *gridpoints*), such that the line-segments representing the edges are pairwise non-crossing. That is, distinct edges only intersect at common endpoints, and each edge only intersects a vertex that is an endpoint of that edge. The *bounding box* of a 3D drawing is the minimum axis-aligned box containing the drawing. If the bounding box has side lengths $X - 1$, $Y-1$ and $Z-1$, then we speak of an $X \times Y \times Z$ drawing with *volume* $X \cdot Y \cdot Z$. That is, the volume of a 3D drawing is the number of gridpoints in the bounding box. Minimising the volume in 3D drawings is a widely studied problem [15, 17, 18, 19, 21, 22, 25, 26, 28, 29, 33, 39, 48, 53]. The following general bounds are known for the volume of 3D drawings in terms of the track-number. Other papers to employ track layouts in the production of 3D drawings include [19, 22, 29, 33, 39].

Table 1: Upper bounds on the track-number.

| graph family | track-number | reference |
| --- | --- | --- |
| $n$ vertices | $n$ | trivial |
| $m$ edges | $15m^{2/3}$ | Dujmović and Wood [29] |
| $m$ edges, max. degree $\Delta$ | $14\sqrt{\Delta m}$ | Dujmović and Wood [29] |
| no $K_h$-minor | $\mathcal{O}(h^{3/2}n^{1/2})$ | Dujmović and Wood [29] |
| genus $\gamma$ | $\mathcal{O}(\gamma^{1/2}n^{1/2})$ | Dujmović and Wood [29] |
| tree-width $w$ | $3^w \cdot 6^{(4^w-3w-1)/9}$ | Dujmović et al. [25, 28] |
| tree-width $w$, max. degree $\Delta$ | $72\Delta w$ | Dujmović et al. [25, 28] |
| queue-number $k$, acyclic chromatic number $c$ | $c(2k)^{c-1}$ | Dujmović et al. [25]; see Theorem 2 |
| queue-number $k$ | $4k \cdot 4^{k(2k-1)(4k-1)}$ | Theorem 8 |
| path-width $p$ | $p+1$ | Dujmović et al. [25, 26] |
| band-width $b$ | $b+1$ | Lemma 19 |
| series-parallel graphs | 15 | Di Giacomo et al. [20] |
| Halin | $8^†$ | Di Giacomo and Meijer [22] |
| X-trees | $6^†$ | Di Giacomo and Meijer [22] |
| outerplanar | $5^†$ | Lemma 22 |
| 1-queue graphs | 4 | Theorem 11 |
| trees | 3 | Felsner et al. [33] |

**Theorem 1.** [25, 26, 29] *Let $G$ be a $c$-colourable $t$-track graph with $n$ vertices. Then*

(a) *$G$ has a $\mathcal{O}(t) \times \mathcal{O}(t) \times \mathcal{O}(n)$ straight-line drawing with $\mathcal{O}(t^2 n)$ volume, and*

(b) *$G$ has a $\mathcal{O}(c) \times \mathcal{O}(c^2 t) \times \mathcal{O}(c^4 n)$ straight-line drawing with $\mathcal{O}(c^7 t n)$ volume.*

*Moreover, if $G$ has an $X \times Y \times Z$ straight-line drawing then $G$ has track-number $\mathsf{tn}(G) \leq 2XY$.*

The purpose of this paper is to present fundamental results in the theory of track layouts. In Section 2.1 we show how to colour the edges of a track assignment to obtain a track layout. In Section 2.2 we answer the extremal question: what is the maximum number of edges in $(k,t)$-track layout? Section 3 presents methods for manipulating track layouts. In particular, we show how to 'wrap' a track layout. This process can be used to produce a track layout of a graph given track layouts of its biconnected components. Section 4 studies the tradeoff between the number of tracks and the number of edge colours in a track layout. In Sections 5 and 6 we explore the relationship between track layouts and other models of graph layout; in particular, stack and queue layouts in Section 5, and geometric thickness in Section 6. One of our main results is that track-number is tied to queue-number. As corollaries we prove that acyclic chromatic number is bounded by both stack-number and queue-number. In Section 7 we prove bounds on the track-number of outerplanar graphs, and prove the best known lower bound on the track-number of planar graphs. Section 8 concludes with some open problems regarding the computational complexity of recognising $(k,t)$-track graphs. Note that a number of results in this paper are used to prove results in our companion paper on layouts of graph subdivisions [27].

---
$^†$A track layout that allows edges between consecutive vertices in a track is called an *improper track layout* [25]. This concept, in the case of three tracks, was introduced by Felsner et al. [33], who proved that every outerplanar graph has an improper 3-track layout. It is easily seen that the tracks can be 'doubled' to obtain a (proper) 6-track layout [25]. Lemma 22 improves this bound to 5. Similarly, Di Giacomo and Meijer [22] proved that X-trees have improper 3-track layouts, and Halin graphs have improper 4-track layouts. Thus X-trees have (proper) 6-track layouts, and Halin graphs have (proper) 8-track layouts.

## 1.1 Definitions

Before we move on, here are some definitions. The subgraph of a graph $G$ induced by a set of vertices $A \subseteq V(G)$ is denoted by $G[A]$. For all disjoint $A, B \subseteq V(G)$, we denote by $G[A, B]$ the bipartite subgraph of $G$ with vertex set $A \cup B$ and edge set $\{vw \in E(G) : v \in A, w \in B\}$. The spanning subgraph of $G$ induced by a set of edges $S \subseteq E(G)$ is denoted by $G[S]$. For $v, w \in V(G)$, we denote by $G \cup vw$ the graph with vertex set $V(G)$ and edge set $E(G) \cup \{vw\}$.

A graph $H$ is a *minor* of $G$ if $H$ is isomorphic to a graph obtained from a subgraph of $G$ by contracting edges. A minor-closed class of graphs is *proper* if it is not the class of all graphs.

A *graph parameter* is a function $\alpha$ that assigns to every graph $G$ a non-negative integer $\alpha(G)$. Let $\mathcal{G}$ be a class of graphs. By $\alpha(\mathcal{G})$ we denote the function $f : \mathbb{N} \to \mathbb{N}$, where $f(n)$ is the maximum of $\alpha(G)$, taken over all $n$-vertex graphs $G \in \mathcal{G}$. We say $\mathcal{G}$ has *bounded* $\alpha$ if $\alpha(\mathcal{G}) \in \mathcal{O}(1)$. A graph parameter $\alpha$ is *bounded by* a graph parameter $\beta$, if there exists a *binding* function $g$ such that $\alpha(G) \leq g(\beta(G))$ for every graph $G$. If $\alpha$ is bounded by $\beta$ and $\beta$ is bounded by $\alpha$ then $\alpha$ and $\beta$ are *tied*. Clearly, if $\alpha$ and $\beta$ are tied then a graph family $\mathcal{G}$ has bounded $\alpha$ if and only if $\mathcal{G}$ has bounded $\beta$. These notions were introduced by Gyárfás [37] in relation to near-perfect graph families for which the chromatic number is bounded by the clique-number.

A *vertex ordering* of an $n$-vertex graph $G$ is a bijection $\sigma : V(G) \to \{1, 2, \ldots, n\}$. We write $v <_\sigma w$ to mean that $\sigma(v) < \sigma(w)$. One can thus view $<_\sigma$ as a total order on $V(G)$. We say $G$ (or $V(G)$) is *ordered by* $<_\sigma$. At times, it will be convenient to express $\sigma$ by the list $(v_1, v_2, \ldots, v_n)$, where $v_i <_\sigma v_j$ if and only if $1 \leq i < j \leq n$. These notions extend to subsets of vertices in the natural way. Suppose that $V_1, V_2, \ldots, V_k$ are disjoint sets of vertices, such that each $V_i$ is ordered by $<_i$. Then $(V_1, V_2, \ldots, V_k)$ denotes the vertex ordering $\sigma$ such that $v <_\sigma w$ whenever $v \in V_i$ and $w \in V_j$ with $i < j$, or $v \in V_i$, $w \in V_i$, and $v <_i w$. We write $V_1 <_\sigma V_2 <_\sigma \cdots <_\sigma V_k$.

## 2 Basics

### 2.1 Fixed Track Assignment

We now show how to colour the edges in a track assignment so that no monochromatic edges form an X-crossing. A set $S$ of $k$ edges in a track assignment $\mathcal{A}$ is called a *crossing $k$-tuple* if each pair of edges in $S$ form an X-crossing in $\mathcal{A}$.

**Lemma 2.** *A $t$-track assignment $\mathcal{A}$ of a graph $G$ can be extended into a $(k,t)$-track layout if and only if $\mathcal{A}$ has no crossing $(k+1)$-tuple.*

*Proof.* Suppose that $\mathcal{A}$ has a crossing $(k+1)$-tuple $S$. Each edge in $S$ must receive a distinct colour. Thus $\mathcal{A}$ cannot be extended into a $(k,t)$-track layout. Now suppose $\mathcal{A}$ has no crossing $(k+1)$-tuple. Consider any two tracks $A, B \in \mathcal{A}$. For all edges $vw, xy$ of $G[A,B]$, say $vw \preceq xy$ if $v \leq x$ in $A$ and $w \leq y$ in $B$. Clearly $\preceq$ is a partial order on $E(G[A,B])$. Two edges are unrelated under $\preceq$ if and only if they form an X-crossing. Thus an antichain in $\preceq$ is a crossing tuple. By assumption, $\preceq$ has no antichain of size $k+1$. By Dilworth's Theorem [24], $E(G[A,B])$ can be partitioned into $k$ chains. A chain in $\preceq$ is a set of edges of $G[A,B]$, no two of which form an X-crossing. Thus the partition into $k$ chains defines the desired edge colouring of a $(k,t)$-track layout. □

Note that Lemma 2 basically says that permutation graphs are perfect.

### 2.2 Extremal Questions

Consider the maximum number of edges in a track layout. It follows from Lemma 1 that an $n$-vertex 2-track graph has at most $n-1$ edges, which generalises to $(k,2)$-track graphs as follows.

**Lemma 3.** *Every $(k,2)$-track graph with $n \geq 1$ vertices has at most $k(n-1)$ edges. Moreover, if $n \geq 2k-1$ (which is implied if in fact there is a crossing $k$-tuple), then there are at most $k(n-k)$ edges. Conversely, for all $k \geq 1$ and $n_1, n_2 \geq k$, there exists a $(k,2)$-track layout with $k(n_1 + n_2 - k)$ edges, and with $n_1$ vertices in the first track and $n_2$ vertices in the second track.*

*Proof.* First we prove the upper bounds. Let $(v_1, v_2, \ldots, v_{n_1})$ and $(w_1, w_2, \ldots, w_{n_2})$ be the tracks, where $n = n_1 + n_2$. For each edge $v_i w_j$, let $\lambda(v_i w_j) = i + j$. Observe that $2 \leq \lambda(v_i w_j) \leq n$. Thus there are at most $n-1$ possible $\lambda$ values. If distinct edges $e$ and $f$ have $\lambda(e) = \lambda(f)$ then $e$ and $f$ form an X-crossing. Thus at most $k$ edges have the same $\lambda$ value. Hence the number of edges is at most $k(n-1)$. Now suppose that $n \geq 2k-1$. For all $1 \leq i \leq k-1$, at most $i$ edges $e$ have $\lambda(e) = i+1$ ($\leq k$), and at most $i$ edges $e$ have $\lambda(e) = n+1-i$ ($\geq k+1$). Thus the number of edges is at most

$$2 \sum_{i=1}^{k-1} i \;+\; \big(n-1-2(k-1)\big)k \;=\; k(n-k) \;.$$

Now we prove the lower bound. Let $A = (v_1, v_2, \ldots, v_{n_1})$ and $B = (w_1, w_2, \ldots, w_{n_2})$. Construct a graph $G$ with $V(G) = A \cup B$. For each $1 \leq \ell \leq k$, let $E_\ell$ be the set of edges

$$\{v_\ell w_j : 1 \leq j \leq n_2 + 1 - \ell\} \bigcup \{v_i w_{n_2 + 1 - \ell} : \ell + 1 \leq i \leq n_1\} \;.$$

Observe that $E_{\ell_1} \cap E_{\ell_2} = \emptyset$ for distinct $\ell_1$ and $\ell_2$. Let $E(G) = \bigcup_\ell E_\ell$. Clearly no two edges in each $E_\ell$ form an X-crossing, as illustrated in Figure 2. Thus $G$ has a $(k, 2)$-track layout. The number of edges is

$$\sum_{\ell=1}^{k} \big((n_2 + 1 - \ell) + (n_1 - \ell)\big) \;=\; k(n_1 + n_2) - \sum_{\ell=1}^{k}(2\ell - 1) \;=\; k(n_1 + n_2 - k) \;.$$

□

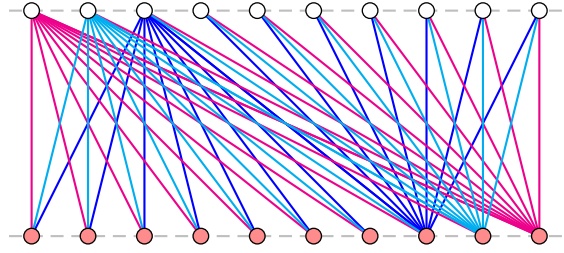

Figure 2: An edge-maximal $(3, 2)$-track layout.

Lemma 3 generalises to $(k, t)$-track layouts as follows.

**Lemma 4.** *Every $(k, t)$-track graph with $n$ vertices and no empty tracks has at most $k\big((t - 1)n - \binom{t}{2}\big)$ edges. Moreover, if every pair of tracks has at least $2k - 1$ vertices, then there are at most $k\big((t-1)n - k\binom{t}{2}\big)$ edges. Conversely, for all $k \geq 1$, $t \geq 2$ and $n \geq kt$ there exists a $(k, t)$-track layout with $n$ vertices and $k\big((t - 1)n - k\binom{t}{2}\big)$ edges.*

*Proof.* First we prove the upper bounds. Let $n_i$ be the number of vertices in the $i$-th track. Let $m_{i,j}$ be the number of edges between the $i$-th and $j$-th tracks. By Lemma 3, $m_{i,j} \leq k(n_i + n_j - 1)$. Hence the total number of edges is at most

$$\sum_{1 \leq i < j \leq t} k(n_i + n_j - 1) \;=\; k\Big(\sum_{1 \leq i < j \leq t}(n_i + n_j) - \binom{t}{2}\Big) \;=\; k\big((t - 1)n - \binom{t}{2}\big) \;.$$

Now suppose that every pair of tracks has at least $2k - 1$ vertices. By Lemma 3, $m_{i,j} \leq k(n_i + n_j - k)$. Hence the total number of edges is at most

$$\sum_{1 \leq i < j \leq t} k(n_i + n_j - k) \;=\; k\Big(\sum_{1 \leq i < j \leq t}(n_i + n_j) - k\binom{t}{2}\Big) \;=\; k\big((t - 1)n - k\binom{t}{2}\big) \;.$$

Now we prove the lower bound. Given any $k \geq 1$, $t \geq 2$ and $n \geq kt$, arbitrarily partition $n$ into $t$ integers $n = n_1 + n_2 + \cdots + n_t$ with each $n_i \geq k$. Construct a $(k, t)$-track layout with $n_i$ vertices in the $i$-th track, and $k(n_i + n_j - k)$ edges between the $i$-th and $j$-th tracks, as in Lemma 3. By the above analysis, the total number of edges is $k\big((t - 1)n - k\binom{t}{2}\big)$. □

Since $\binom{t}{2} \geq 1$, Lemma 4 implies the following lower bound on $\mathsf{tn}_k(G)$.

**Corollary 1.** *For all $k \geq 1$, every graph $G$ with $n$ vertices and $m \geq 1$ edges satisfies $\mathsf{tn}_k(G) \geq \frac{m+k}{kn} + 1$.* □

# 3 Manipulating Track Layouts

## 3.1 The Wrapping Lemma

Consider a track assignment $\{V_i : 1 \leq i \leq t\}$ with a fixed ordering of the tracks. The *span* of an edge $vw$ in $\{V_i : 1 \leq i \leq t\}$ is $|i - j|$ where $v \in V_i$ and $w \in V_j$. It will also be useful to consider track layouts whose index set is two-dimensional. Let $\{V_{i,j} : i \geq 0, 1 \leq j \leq b_i\}$ be a track assignment of a graph $G$. Define the *partial span* of an edge $vw \in E(G)$ with $v \in V_{i_1, j_1}$ and $w \in V_{i_2, j_2}$ to be $|i_1 - i_2|$.

The following lemma describes how to 'wrap' a track layout, and is a generalisation of a result by Dujmović *et al.* [25, 26], which in turn is based on an idea due to Felsner *et al.* [33].

**Lemma 5.** *Let $\{V_{i,j} : i \geq 0, 1 \leq j \leq b_i\}$ be a $(k, t)$-track layout of a graph $G$ with maximum partial span $s$ (for some irrelevant value $t$). For each $0 \leq \alpha \leq s$, let $t_\alpha = \max\{b_i : i \equiv \alpha \pmod{s+1}\}$. For each $0 \leq \alpha \leq 2s$, let $t'_\alpha = \max\{b_i : i \equiv \alpha \pmod{2s+1}\}$. Then*

$$\text{(a) } \mathsf{tn}_{2k}(G) \leq \sum_{\alpha=0}^{s} t_\alpha \ , \ \text{and} \ \text{(b) } \mathsf{tn}_{k}(G) \leq \sum_{\alpha=0}^{2s} t'_\alpha \ .$$

*Proof.* Let $\{E_\ell : 1 \leq \ell \leq k\}$ be the edge colouring in the given track layout. First we prove (a). By adding extra empty tracks where necessary, we can assume that the track layout is indexed

$$\{V_{i,j} : i \geq 0, 1 \leq j \leq t_\alpha, \alpha = i \bmod (s+1)\} \ .$$

For each $0 \leq \alpha \leq s$ and $1 \leq j \leq t_\alpha$, let

$$W_{\alpha,j} = \bigcup \{V_{i,j} : i \equiv \alpha \pmod{s+1}, i \geq 0\} \ .$$

Order $W_{\alpha,j}$ by

$$(V_{\alpha,j},\ V_{\alpha+(s+1),j},\ V_{\alpha+2(s+1),j},\ \dots) \ .$$

Since every edge $vw \in E(G)$ has partial span at most $s$, if $v \in W_{\alpha_1, j_1}$ and $w \in W_{\alpha_2, j_2}$ then $\alpha_1 \neq \alpha_2$ or $j_1 \neq j_2$. Hence $\{W_{\alpha,j} : 0 \leq \alpha \leq s, 1 \leq j \leq t_\alpha\}$ is a track assignment of $G$. For each $1 \leq \ell \leq k$, let

$$E'_\ell = \{vw \in E_\ell : v \in V_{i_1, j_1} \cap W_{\alpha_1, j_1},\ w \in V_{i_2, j_2} \cap W_{\alpha_2, j_2}, i_1 \leq i_2, \alpha_1 \leq \alpha_2\}, \text{ and}$$

$$E''_\ell = \{vw \in E_\ell : v \in V_{i_1, j_1} \cap W_{\alpha_1, j_1},\ w \in V_{i_2, j_2} \cap W_{\alpha_2, j_2}, i_1 < i_2, \alpha_2 < \alpha_1\} \ .$$

An X-crossing between edges both from some $E'_\ell$ (or both from some $E''_\ell$) implies that the same edges form an X-crossing in the original track layout. Thus $\{E'_\ell, E''_\ell : 1 \leq \ell \leq k\}$ defines an edge $2k$-colouring with no monochromatic X-crossing. Thus we have a $(2k, \sum_{\alpha=0}^{s} t_\alpha)$-track layout of $G$.

We now prove (b). Again by adding extra empty tracks where necessary, we can assume that the track layout is indexed $\{V_{i,j} : i \geq 0, 1 \leq j \leq t'_\alpha, \alpha = i \bmod (2s+1)\}$. For each $0 \leq \alpha \leq 2s$ and $1 \leq j \leq t'_\alpha$, let

$$W_{\alpha,j} = \bigcup \{V_{i,j} : i \equiv \alpha \pmod{2s+1}, i \geq 0\} \ .$$

Order $W_{\alpha,j}$ by

$$(V_{\alpha,j},\ V_{\alpha+(2s+1),j},\ V_{\alpha+2(2s+1),j},\ \dots) \ .$$

Clearly $\{W_{\alpha,j} : 0 \leq \alpha \leq 2s, 1 \leq j \leq t'_\alpha\}$ is a track assignment of $G$. It remains to prove that there is no monochromatic X-crossing, where edge colours are inherited from the given track layout. Notice that each $E_\ell = E'_\ell \cup E''_\ell$. As in part (a), edges in $E'_\ell$ or in $E''_\ell$ do not form an X-crossing. In the track layout defined for part (b), edges in $E'_\ell$ have partial span at most $s$, and edges in $E''_\ell$ have partial span at least $s+1$. Thus an edge from $E'_\ell$ and an edge from $E''_\ell$ do not form an X-crossing. Hence we have a $(k, \sum_{\alpha=0}^{2s} t'_\alpha)$-track layout of $G$. □

The full generality of Lemma 5 is used in our companion paper [27]. For other applications, the following two special cases suffice. By Lemma 5 with $b_i = b$ for all $i \geq 0$, we have:

**Lemma 6.** *Let $\{V_{i,j} : i \geq 0, 1 \leq j \leq b\}$ be a $(k,t)$-track layout of a graph $G$ with maximum partial span $s$. Then (a) $\mathsf{tn}_{2k}(G) \leq (s+1)b$, and (b) $\mathsf{tn}_k(G) \leq (2s+1)b$.* □

The next special case is Lemma 6 with $b = 1$. Lemma 7(b) with $k = 1$ was proved by Dujmović *et al.* [25, 26].

**Lemma 7.** *Let $G$ be a $(k,t)$-track graph with maximum span $s$. Then (a) $\mathsf{tn}_{2k}(G) \leq s+1$, and (b) $\mathsf{tn}_k(G) \leq 2s+1$.* □

## 3.2 Biconnected Components

Clearly the track-number of a graph is at most the maximum track-number of its connected components. We now prove a similar result for maximal biconnected components (*blocks*).

**Lemma 8.** *For every $k \geq 1$, every graph $G$ satisfies:*

(a) $\mathsf{tn}_{2k}(G) \leq 2 \cdot \max\{\mathsf{tn}_k(B) : B \text{ is a block of } G\}$, *and*

(b) $\mathsf{tn}_k(G) \leq 3 \cdot \max\{\mathsf{tn}_k(B) : B \text{ is a block of } G\}$.

*Proof.* Suppose we have a $(k,t)$-track layout of each block of $G$, where $t = \max\{\mathsf{tn}_k(B) : B \text{ is a block of } G\}$. Clearly we can assume that $G$ is connected. Let $T$ be the *block-cut-tree* of $G$. That is, there is one vertex in $T$ for each block and for each cut-vertex of $G$. Two vertices of $T$ are adjacent if one corresponds to a block $B$ and the other corresponds to a cut-vertex $v \in B$. $T$ is a tree, as otherwise a cycle in $T$ would correspond to a single block of $G$. Root $T$ at a node $r$ corresponding to an arbitrary block.

A node of $T$ that corresponds to a block of $G$ is at even distance from $r$, and a node of $T$ that corresponds to a cut-vertex of $G$ is at odd distance from $r$. For all $i \geq 0$, let $D_i$ be the set of blocks of $G$ whose corresponding node in $T$ is at distance $2i$ from $r$. Consider a block $B \in D_i$. Let $x$ be the node of $T$ corresponding to $B$. Let $p$ be the parent node of $x$ in $T$. Then $p$ corresponds to a cut-vertex of $G$, which we call the *parent cut-vertex* of $B$. Say $i \geq 1$. Let $y$ be the parent node of $p$ in $T$. Then $y$ corresponds to some block $B'$ of $G$. We say $B'$ is the *parent block* of $B$, and $B$ is a *child block* of $B'$. Observe that each cut-vertex $v$ is the parent cut-vertex of all but one block containing $v$. If a vertex $v$ of $G$ is in only one block $B$ then we say $v$ is *grouped* with $B$. Otherwise $v$ is a cut-vertex and we say $v$ is *grouped* with the block for which it is not the parent block.

Now order each $D_i$ firstly with respect to the order of the parent blocks in $D_{i-1}$, and secondly with respect to the order of the parent cut-vertices in the track layouts of the parent blocks. More formally, for each $i \geq 1$, let $<_i$ be a total order of $D_i$ such that for all blocks $A, B \in D_i$ (with parent blocks $A', B' \in D_{i-1}$) we have $A <_i B$ whenever (1) $A' <_{i-1} B'$, or (2) $A' = B'$, $A \cap A' = \{v\}$, $B \cap B' = \{w\}$, and $v < w$ in some track of the $(k,t)$-track layout of $A'$. (If $v$ and $w$ are in different tracks of the $(k,t)$-track layout of the parent block then the relative order of $A$ and $B$ is not important.)

For each $i \geq 0$ and $1 \leq j \leq t$, let $V_{i,j}$ be the set of vertices $v$ of $G$ in a some block $B \in D_i$ such that $v$ is grouped with $B$, and $v$ is in the $j$-th track of the track layout of $B$. Now order each $V_{i,j}$ firstly with respect to the order $<_i$ of the blocks in $D_i$, and within a block $B$, by the order of the $j$-th track of the track layout of $B$. Colour each edge $e$ of $G$ by the same colour assigned to $e$ in the $(k,t)$-track layout of the block containing $e$. We claim there is no monochromatic X-crossing.

The parent cut-vertex of a block $B$ is grouped with the parent block of $B$, and no block and its parent block are in the same $D_i$. Thus if $vw$ is an edge with $v \in V_{i,j_1}$ and $w \in V_{i,j_2}$ then both $v$ and $w$ are grouped with the block containing $vw$. Since within each track vertices are ordered primarily by their block, and by assumption there is no monochromatic X-crossing between edges in the same block, there is no monochromatic X-crossing between tracks $V_{i,j_1}$ and $V_{i,j_2}$ for all $i \geq 0$ and $1 \leq j_1, j_2 \leq t$.

If $vw$ is an edge with $v \in V_{i_1, j_1}$ and $w \in V_{i_2, j_2}$ for distinct $i_1$ and $i_2$, then without loss of generality, $i_2 = i_1 + 1$ and $v$ is the parent cut-vertex of the block containing $vw$. Since sibling blocks are ordered with respect to the ordering of their parent cut-vertices, there is no X-crossing amongst edges between tracks $V_{i_1, j_1}$ and $V_{i_2, j_2}$ for all $i_1, i_2 \geq 0$ and $1 \leq j_1, j_2 \leq t$. Thus $\{V_{i,j} : i \geq 0, 1 \leq j \leq t\}$ is a $k$-edge colour track layout of $G$ such that every edge has a partial span of one. By Lemma 6, $G$ has $\mathsf{tn}_{2k}(G) \leq 2t$, and $G$ has $\mathsf{tn}_k(G) \leq 3t$. □

## 4 Tracks vs. Colours

We now show how to reduce the number of tracks in a track layout, at the expense of increasing the number of edge colours.

**Lemma 9.** *Let $G$ be a $(k,t)$-track graph with maximum span $s$ ($\leq t-1$). For every vertex colouring $\{V_i : 1 \leq i \leq c\}$ of $G$, there is a $(2sk, c)$-track layout of $G$ with tracks $\{V_i : 1 \leq i \leq c\}$.*

*Proof.* Let $\{T_j : 1 \leq j \leq t\}$ be a $(k,t)$-track layout of $G$ with maximum span $s$ and edge colouring $\{E_\ell : 1 \leq \ell \leq k\}$. Given a vertex colouring $\{V_i : 1 \leq i \leq c\}$ of $G$, order each $V_i$ by $(V_i \cap T_1, V_i \cap T_2, \ldots, V_i \cap T_t)$. Thus $\{V_i : 1 \leq i \leq c\}$ is a $c$-track assignment of $G$. Now we define an edge $2sk$-colouring. For each $\ell$ and $\alpha$ such that $1 \leq \ell \leq k$ and $1 \leq |\alpha| \leq s$, let

$$E_{\ell, \alpha} = \{vw \in E_\ell : v \in V_{i_1} \cap T_{j_1}, w \in V_{i_2} \cap T_{j_2}, i_1 < i_2, j_1 - j_2 = \alpha\} \ .$$

Consider two edges $vw$ and $xy$ in some $E_{\ell, \alpha}$ between a pair of tracks $V_{i_1}$ and $V_{i_2}$. Without loss of generality $i_1 < i_2$, $v \in V_{i_1} \cap T_{j_1}$, $w \in V_{i_2} \cap T_{j_1 + \alpha}$, $x \in V_{i_1} \cap T_{j_2}$, $y \in V_{i_2} \cap T_{j_2 + \alpha}$, and $j_1 \leq j_2$. If $j_1 = j_2$ then $vw$ and $xy$ are between the same pair of tracks in the given track layout, and the relative order of the vertices is preserved. Thus if $vw$ and $xy$ form an X-crossing in the $c$-track assignment then they are coloured differently. If $j_1 < j_2$ then $v <_{i_1} x$ and $w <_{i_2} y$, and the edges do not form an X-crossing. Hence $vw$ and $xy$ do not form a monochromatic X-crossing, and we have a $(2sk, c)$-track layout of $G$. □

We now show how to reduce the number of edge colours in a track layout, at the expense of increasing the number of tracks. A vertex colouring is *acyclic* if there is no bichromatic cycle; that is, every cycle receives at least three colours. The *acyclic chromatic number* of a graph $G$, denoted by $\chi_a(G)$, is the minimum number of colours in an acyclic vertex colouring of $G$. This concept was introduced by Grünbaum [36], and has since been widely studied [3, 4, 5, 10, 11, 12, 13, 14, 14, 16, 34, 35, 43, 47]. By Lemma 1, each 2-track subgraph in an (edge-monochromatic) track layout is a forest of caterpillars. Thus the underlying vertex colouring is acyclic. Hence,

$$\chi_a(G) \leq \mathsf{tn}(G) \ . \tag{1}$$

A number of the results in this paper bound the acyclic chromatic number by various 'geometric' graph parameters. Many other variations of the chromatic number (including star chromatic number [1, 35, 36] and oriented chromatic number [44, 50, 52]) are bounded by the acyclic chromatic number. Thus our results also apply to these other types of colourings—we omit the details.

Alon and Marshall [2] proved the following application of acyclic colourings that we will repeatedly use. Loosely speaking, it says how to transform an edge colouring into a vertex colouring. A vertex colouring $C_1$ is a *refinement* of a vertex colouring $C_2$ if every colour class of $C_1$ is a subset of some colour class of $C_2$.

**Lemma 10.** [2] *Given an edge $k$-colouring of a graph $G$, any acyclic $c$-colouring of $G$ can be refined to a $ck^{c-1}$-colouring so that the edges between any pair of (vertex) colour classes are monochromatic.*

The following result, which is implicit in Lemma 5.3 of Dujmović *et al.* [25], easily follows from Lemma 10.

**Theorem 2.** [25] *Let $G$ be a $(k,t)$-track graph in which the underlying vertex $t$-colouring is acyclic. Then $G$ has track-number $\mathsf{tn}(G) \leq tk^{t-1}$.*

The following result is similar to Theorem 2 but without the assumption that $G$ has an acyclic colouring. A track layout $T_1$ is a *refinement* of a track layout $T_2$ if every track in $T_1$ is a subset of, and has the same order as, some track in $T_2$.

**Theorem 3.** *Every $(k,t)$-track layout of a graph $G$ can be refined to an (edge-monochromatic) $t \cdot 4^{\binom{k}{2}(t-1)}$-track layout of $G$. That is, $G$ has track-number $\mathsf{tn}(G) \leq t \cdot 4^{\binom{k}{2}(t-1)}$.*

We will prove Theorem 3 by a series of lemmas. Recall that a *crossing triple* in a track assignment is a set of three edges, each pair of which forms an X-crossing. The following result is the key idea.

**Lemma 11.** *Every 2-track assignment with no crossing triple can be refined to an (edge-monochromatic) 8-track layout, with four tracks arising from each of the two given tracks.*

*Proof.* Let $\{A, B\}$ be a 2-track assignment of a graph $G$ with no crossing triple. We consider each track to be ordered left-to-right. Construct a path $P$ starting at the first vertex in $A$ as follows. If $v$ is the current endpoint of $P$, choose $vw$ to be the next edge in $P$, where $w$ is the rightmost vertex in the track that does not contain $v$, such that $G \cup vw$ has no crossing triple. (Note that $vw$ is not necessarily an edge of $G$.) Repeat this process until $v$ is the rightmost vertex in $B$. Then $\{A, B\}$ is a 2-track assignment of $G \cup P$ with no crossing triple.

In the construction of $P$, we can always choose the edge $vw$, since if $uv$ is the edge most recently added to $P$, then the first vertex $w$ to the right of $u$ satisfies the conditions on $w$ (since if $\{vw, e_1, e_2\}$ is a crossing triple, then $\{uv, e_1, e_2\}$ is a crossing triple, or $v$ was not chosen rightmost.) This also proves that $P$ is plane (that is, non-self-crossing). Moreover, $P$ is an induced path, as otherwise some vertex in $P$ would not be rightmost.

**Claim.** Each edge of $G$ crosses at most one edge of $P$.

*Proof.* Assume for the sake of contradiction that an edge $xy$ of $G$ crosses two edges of $P$. Then $xy$ crosses some 2-edge path $uvw \in P$, such that without loss of generality $u \in B$, $v \in A$, $w \in B$, and $u < w$ in $B$. We can assume that $x$ is the rightmost vertex in its track that has an incident edge $xy$ that crosses $uvw$, and $x < v$ in $A$ or $x < u$ in $B$.

Case 1. $x \in A$ (see Figure 3(a)): Since $vw$ was added to $P$ (and $vy$ was not added to $P$), there are two edges $pq, rs \in G$ ($p, r \in A$ and $q, s \in B$), such that $\{pq, rs, vy\}$ is a crossing triple. If $pq$ or $rs$ does not cross $xy$, then $x < p < v$ or $x < r < v$, in which case $pq$ or $rs$ crosses $uvw$ with $p > x$ or $r > x$, which contradicts our choice of $x$. Otherwise $\{pq, rs, xy\}$ is a crossing triple, which is again a contradiction.

Case 2. $x \in B$ (see Figure 3(b)): Since $uv$ was added to $P$ (and $uy$ was not added to $P$), there are two edges $pq, rs \in G$ ($p, r \in A$ and $q, s \in B$), such that $\{pq, rs, uy\}$ is a crossing triple. If $pq$ or $rs$ does not cross $xy$ then $x < q < u$ or $x < s < u$, in which case $pq$ or $rs$ crosses $uvw$ with $x < q$ or $x < s$, which contradicts our choice of $x$. Otherwise $\{pq, rs, xy\}$ is a crossing triple, which is again a contradiction. □

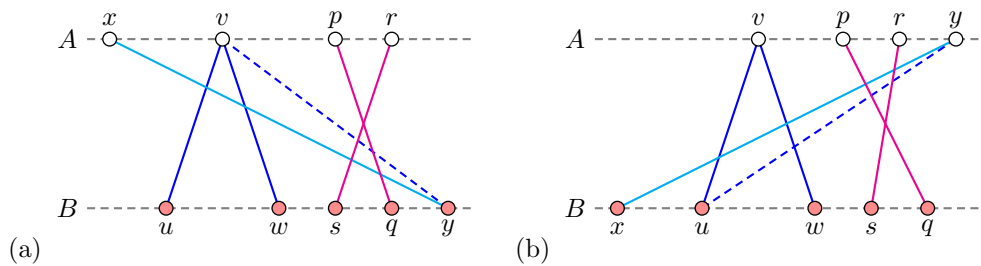

Figure 3: Illustration to show that every edge of $G$ crosses at most one edge of $P$.

Let $P = (v_1, w_1, v_2, w_2, \ldots, v_t, w_t)$, where $v_1$ is the first vertex in $A$, and $v_1 < v_2 < \cdots < v_t$ in $A$, and $w_t$ is the last vertex in $B$, and $w_1 < w_2 < \cdots < w_t$ in $B$.

For $1 \le i \le t - 1$, let $A_i$ be the subset of $A$ consisting of the vertices strictly between $v_i$ and $v_{i+1}$. Let $A_t$ be the subset of $A$ consisting of the vertices strictly after $v_t$. Let $B_0$ be the subset of $B$ consisting of the vertices strictly before $w_1$. For $1 \le i \le t - 1$, let $B_i$ be the subset of $B$ consisting of the vertices strictly between $w_i$ and $w_{i+1}$.

Let $X_1 = \{v_i : i \text{ odd}\}$, $X_2 = \{v_i : i \text{ even}\}$, $X_3 = \cup\{A_i : i \text{ odd}\}$, and $X_4 = \cup\{A_i : i \text{ even}\}$.

Let $Y_1 = \{w_i : i \text{ odd}\}$, $Y_2 = \{w_i : i \text{ even}\}$, $Y_3 = \cup\{B_i : i \text{ odd}\}$, and $Y_4 = \cup\{B_i : i \text{ even}\}$.

Consider each of $X_1$, $X_2$, $X_3$ and $X_4$ to be tracks ordered as in $A$. Consider each of $Y_1$, $Y_2$, $Y_3$ and $Y_4$ to be tracks ordered as in $B$. Note that every vertex of $G$ is in one of these tracks. We claim that there is no X-crossing between these tracks.

Since $P$ is an induced path of $G\cup P$, every edge of $G[X_1\cup X_2, Y_1\cup Y_2]$ is in $P$. Since $P$ is non-crossing, there is no X-crossing between tracks $X_\alpha$ and $Y_\beta$ for all $\alpha,\beta \in \{1,2\}$.

Consider the subgraphs $G[X_1, Y_3\cup Y_4]$ and $G[X_2, Y_3\cup Y_4]$. By the construction of $P$, $G[X_1, Y_3\cup Y_4]$ only has edges $v_i x$ where $i$ is odd and $x\in B_{i-1}\cup B_{i-2}$, and $G[X_2, Y_3\cup Y_4]$ only has edges $v_i x$ where $i$ is even and $x\in B_{i-1}\cup B_{i-2}$. Thus $G[X_1, Y_3\cup Y_4]$ and $G[X_2, Y_3\cup Y_4]$ consist of non-crossing stars rooted at vertices $v_i$ of $P$. Similarly, $G[Y_1, X_3\cup X_4]$ and $G[Y_2, X_3\cup X_4]$ consist of non-crossing stars rooted at vertices $w_i \in P$. Thus there is no X-crossing in $G[X_1, Y_3\cup Y_4]$, $G[X_2, Y_3\cup Y_4]$, $G[Y_1, X_3\cup X_4]$ and $G[Y_2, X_3\cup X_4]$.

Now, assume for the sake of contradiction that two edges $vw$ and $xy$, in $G[X_3, Y_3]$ form an X-crossing. Say $v<x$ in $A$ and $w<y$ in $B$. Since no edge of $G[X_3, Y_3]$ crosses two edges of $P$, we have $v,x \in A_i$ and $w,y \in B_i$ for some $i$. Thus $v<x<v_{i+1}$ and $w_i<y<w$. Hence $\{vw, xy, w_i v_{i+1}\}$ is a crossing triple, which is a contradiction. It is simple to verify that the same arguments prove that there is no X-crossing in $G[X_3, Y_4]$, $G[X_4, Y_3]$ and $G[X_4, Y_4]$.

This completes the proof that $\{X_1, X_2, X_3, X_4, Y_1, Y_2, Y_3, Y_4\}$ is the desired 8-track layout of $G$. □

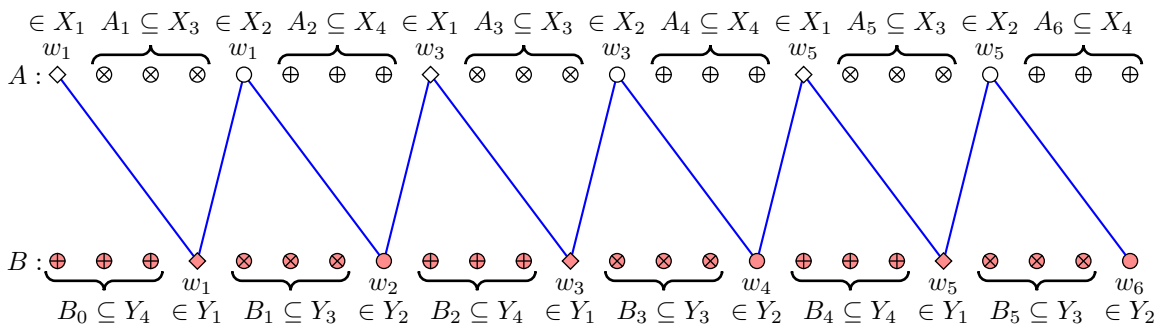

Figure 4: Construction of an 8-track refinement of a $(2,2)$-track layout.

**Lemma 12.** *Every $(k,2)$-track layout can be refined to an (edge-monochromatic) $2\cdot 4^{\binom{k}{2}}$-track layout, with $4^{\binom{k}{2}}$ tracks arising from each of the two given tracks.*

*Proof.* Let $\{A, B\}$ be the tracks and let $\{E_1, E_2, \ldots, E_k\}$ be the edge colouring in a $(k,2)$-track layout of a graph $G$. By Lemma 11, for each pair $\{i,j\} \in \binom{[k]}{2}$, there is an 8-track layout $\{X^{i,j}_\alpha, Y^{i,j}_\alpha : 1\leq \alpha \leq 4\}$ of $G[E_i\cup E_j]$, where every $X^{i,j}_\alpha \subseteq A$ and $Y^{i,j}_\alpha \subseteq B$. For each vertex $v\in A$, define the vector $(X^{i,j}_\alpha : \{i,j\} \in \binom{[k]}{2}, v\in X^{i,j}_\alpha)$. For each vertex $w\in B$, define the vector $(Y^{i,j}_\alpha : \{i,j\} \in \binom{[k]}{2}, w\in Y^{i,j}_\alpha)$. Now group the vertices with a common vector into a track, ordered by $A$ or $B$ accordingly. Since each of the 8-track layouts is a refinement of $\{A,B\}$, the order of the vertices within each of the 8-track layouts is preserved. If two edges coloured $i$ form an X-crossing, then the same pair of edges would form an X-crossing in the $(k,2)$-track layout of $G$. If two edges coloured $i$ and $j$ form an X-crossing, then the same edges would form an X-crossing in the 8-track layout of $G[E_i\cup E_j]$. Hence there is no X-crossing. The number tracks for each of $A$ and $B$ is $4^{\binom{k}{2}}$. □

*Proof of Theorem 3.* Let $\{V_1, V_2, \ldots, V_t\}$ be the tracks in a $(k,t)$-track layout of $G$. For each vertex $v\in V_i$, and for every other track $V_j$, let $\phi_j(v)$ be the track containing $v$ in the track layout of $G[V_i, V_j]$ determined by Lemma 12, where $1 \leq \phi_j(v) \leq 4^{\binom{k}{2}}$. For each vertex $v \in V_i$, define the vector $(i; \phi_1(v), \ldots, \phi_{i-1}(v), \phi_{i+1}(v), \ldots, \phi_t(v))$. Group the vertices with a common vector into a track, ordered by the appropriate $V_i$. Since each track layout of $G[V_i, V_j]$ is a refinement of $\{V_i, V_j\}$, the order of the vertices within each track layout of $G[V_i, V_j]$ is preserved. If two edges are between the same pair of tracks, then their endpoints must be from the same pair of colour classes. Thus there is no X-crossing, as otherwise there would be an X-crossing in the track layout of some $G[V_i, V_j]$. The total number of tracks is $t \cdot 4^{\binom{k}{2}(t-1)}$. □

Note that Lemma 12 can re-stated as follows:

**Corollary 2.** *The vertices of a $(k, 2)$-track graph can be coloured with $2 \cdot 4^{\binom{k}{2}}$ colours so that each bichromatic subgraph is a plane caterpillar; that is, each bichromatic subgraph has no X-crossing.* □

Theorem 3 and (1) imply:

**Corollary 3.** *Every $(k, t)$-track graph $G$ has acyclic chromatic number $\chi_a(G) \leq t \cdot 4^{\binom{k}{2}(t-1)}$.* □

Note that the converse of Corollary 3 is not true. Let $K_n''$ be the graph obtained from the complete graph $K_n$ by subdividing every edge twice. Colour each original vertex of $K_n$ red, and colour the two division vertices of each edge blue and green. Clearly there is no bichromatic cycle. Thus $\chi_a(K_n'') = 3$ for $n \geq 3$. However, $K_n''$ has track-number $\mathsf{tn}(K_n'') \in \Omega(n^{1/4})$ [27]. Thus track-number is not bounded by acyclic chromatic number.

## 5 Queue and Stack Layouts

In a vertex ordering $\sigma$ of a graph $G$, let $L(e)$ and $R(e)$ denote the endpoints of each edge $e \in E(G)$ such that $L(e) <_\sigma R(e)$. Consider two edges $e, f \in E(G)$ with no common endpoint such that $L(e) <_\sigma L(f)$. If $L(e) <_\sigma L(f) <_\sigma R(e) <_\sigma R(f)$ then $e$ and $f$ *cross*, and if $L(e) <_\sigma L(f) <_\sigma R(f) <_\sigma R(e)$ then $e$ and $f$ *nest*, and $f$ is *nested inside* $e$. A *stack* (respectively, *queue*) is a set of edges $E' \subseteq E(G)$ such that no two edges in $E'$ cross (nest). A $k$-*stack* (*queue*) *layout* of $G$ consists of a vertex ordering $\sigma$ of $G$ and a partition $\{E_\ell : 1 \leq \ell \leq k\}$ of $E(G)$, such that each $E_\ell$ is a stack (queue) in $\sigma$. A graph admitting a $k$-stack (queue) layout is called a $k$-*stack* (*queue*) *graph*. The *stack-number* of a graph $G$, denoted by $\mathsf{sn}(G)$, is the minimum $k$ such that $G$ is a $k$-stack graph. Note that stack-number is also called *page-number* and *book-thickness*. The *queue-number* of a graph $G$, denoted by $\mathsf{qn}(G)$, is the minimum $k$ such that $G$ is a $k$-queue graph. See our companion paper [30] for a list of references and applications of stack and queue layouts.

Bernhart and Kainen [6] observed that the 1-stack graphs are precisely the outerplanar graphs, and that 2-stack graphs are characterised as the subgraphs of planar Hamiltonian graphs. Heath and Rosenberg [41] characterised 1-queue graphs as the 'arched levelled planar' graphs.

The following lemma highlights the fundamental relationship between track layouts, and queue and stack layouts. Its proof follows immediately from the definitions, and is illustrated in Figure 5 for $k = 1$.

**Lemma 13.** *Let $\{A, B\}$ be a track assignment of a bipartite graph $G$. Then the following are equivalent:*

(a) *$\{A, B\}$ admits a $(k, 2)$-track layout of $G$,*

(b) *the vertex ordering $(A, B)$ admits a $k$-queue layout of $G$, and*

(c) *the vertex ordering $(A, \overleftarrow{B})$ admits a $k$-stack layout of $G$,*

*where $\overleftarrow{B}$ denotes the reverse vertex ordering of $B$.* □

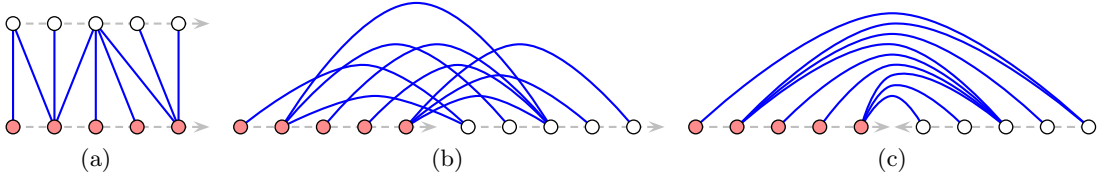

Figure 5: Layouts of a caterpillar: (a) 2-track, (b) 1-queue, (c) 1-stack.

Lemma 13 implies that a $(k, 2)$-track layout can be thought of as a $k$-stack layout of a bipartite graph in which the two colour classes are 'separated' in the ordering. In Corollary 2 we proved that the vertices of $(k, 2)$-track layout can be coloured with $2 \cdot 4^{\binom{k}{2}}$ colours so that each bichromatic subgraph is crossing-free. This result can be generalised as follows.

**Theorem 4.** *Every $k$-stack graph has a vertex $80^{\binom{k}{2}}$-colouring so that each bichromatic subgraph is contained in a single stack, and is thus crossing-free and outerplanar.*

*Proof.* Every 2-stack graph is planar [6], and thus has an acyclic 5-colouring [10]. By Lemma 10 using the stack assignment as an edge 2-colouring, every 2-stack graph has a vertex colouring with $5 \cdot 2^4 = 80$ colours, so that the edges of each bichromatic subgraph are in a single stack.

Now suppose that $G$ is a $k$-stack graph. Let $\{E_1, E_2, \ldots, E_k\}$ be the stacks. For each pair $\{i, j\} \in \binom{[k]}{2}$, there is a vertex 80-colouring of $G[E_i \cup E_j]$, such that the edges in each bichromatic subgraph are contained in a single stack. Let $\phi_{i,j}(v) \in \{1, 2, \ldots, 80\}$ be the colour assigned to each vertex $v$. Now colour $v$ by the vector $\phi(v) = (\phi_{i,j}(v) : \{i, j\} \in \binom{[k]}{2})$.

Suppose there are two edges $vw$ and $xy$ in some bichromatic subgraph of $\phi$, and $vw$ and $xy$ are in distinct stacks $E_i$ and $E_j$. Without loss of generality, $\phi(v) = \phi(x)$ and $\phi(w) = \phi(y)$. Thus $\phi_{i,j}(v) = \phi_{i,j}(x)$ and $\phi_{i,j}(w) = \phi_{i,j}(y)$. Hence in the vertex 80-colouring of $G[E_i \cup E_j]$, there is a bichromatic subgraph with two edges in distinct stacks, which is a contradiction. Thus each bichromatic subgraph of $\phi$ is contained in a single stack. The number of colours is $80^{\binom{k}{2}}$. □

**Theorem 5.** *Acyclic chromatic number is bounded by stack-number. In particular, every $k$-stack graph $G$ has acyclic chromatic number $\chi_a(G) \leq 80^{k(2k-1)}$.* □

*Proof.* The edges of an outerplanar graph can be partitioned into two acyclic subgraphs [46]. Thus $G$ has a $2k$-stack layout in which each stack is acyclic. The result follows from Theorem 4. □

Note that the converse of Theorem 5 is not true. Blankenship and Oporowski [7, 8, 9] proved that the stack-number of $K_n''$ is unbounded, but $\chi_a(K_n'') = 3$ for $n \geq 3$ (see Section 4). Thus stack-number is not bounded by acyclic chromatic number.

## 5.1 Tracks into Queues

Consider how to convert a track layout into a queue layout. The following lemma was proved by Dujmović *et al.* [25, 53] for $k=1$, and with $t=2$ is nothing more than Lemma 13(b).

**Lemma 14.** *Queue-number is bounded by track-number. In particular, every $(k,t)$-track graph with maximum span $s$ ($\leq t-1$) has a $ks$-queue layout.*

*Proof.* Let $\{V_i : 1 \leq i \leq t\}$ be a $(k,t)$-track layout of a graph $G$ with maximum span $s$ and edge colouring $\{E_\ell : 1 \leq \ell \leq k\}$. Let $\sigma$ be the vertex ordering $(V_1, V_2, \ldots, V_t)$ of $G$. Let $E_{\ell,\alpha}$ be the set of edges in $E_\ell$ with span $\alpha$ in the given track layout. Two edges from the same pair of tracks are nested in $\sigma$ if and only if they form an X-crossing in the track layout. Since no two edges in $E_\ell$ form an X-crossing in the track layout, no two edges in $E_\ell$ and between the same pair of tracks are nested in $\sigma$. If two edges not from the same pair of tracks have the same span then they are not nested in $\sigma$. (This idea is due to Heath and Rosenberg [41].) Thus no two edges are nested in each $E_{\ell,\alpha}$, and we have a $ks$-queue layout of $G$. □

Lemmata 9 and 13 imply an analogous result to Lemma 14 for stack layouts of bipartite graphs.

**Lemma 15.** *Every bipartite $(k,t)$-track graph with maximum span $s$ ($\leq t-1$) has a $2ks$-stack layout.* □

## 5.2 Queues into Tracks

Now consider how to convert a vertex ordering into a track layout. The proof of the next result follows immediately from the definitions, and is illustrated in Figure 6.

**Lemma 16.** *Let $\sigma$ be a vertex ordering of a graph $G$. Let $\{V_i : 1 \leq i \leq c\}$ be a vertex colouring of $G$. For all $1 \leq i, j \leq c$, a pair of edges $vw, xy \in E(G[V_i, V_j])$ form an X-crossing in the track assignment $\{(V_i, \sigma) : 1 \leq i \leq c\}$ if and only if:*

- *$vw$ and $xy$ are nested in $\sigma$ (Figures 6(a)-(b)), or*

- *$vw$ and $xy$ cross in $\sigma$ with $v <_\sigma y <_\sigma w <_\sigma x$, and $v, x \in V_i$ and $w, y \in V_j$ (Figure 6(e)).* □

The following observation is implicit in Lemma 5.3 of Dujmović *et al.* [25].

**Lemma 17.** *For every vertex colouring $\{V_i : 1 \leq i \leq c\}$ of a $q$-queue graph $G$, there is a $(2q, c)$-track layout of $G$ with tracks $\{V_i : 1 \leq i \leq c\}$.*

*Proof.* Let $\sigma$ be the vertex ordering in a $q$-queue layout of $G$ with queues $\{E_\ell : 1 \leq \ell \leq q\}$. Let $\{(V_i, \sigma) : 1 \leq i \leq c\}$ be a $c$-track assignment of $G$, and for each $1 \leq \ell \leq q$, let

$$E'_\ell = \{vw \in E_\ell : v \in V_i, w \in V_j, i < j, v <_\sigma w\}, \text{ and}$$
$$E''_\ell = \{vw \in E_\ell : v \in V_i, w \in V_j, i < j, w <_\sigma v\} \ .$$

By Lemma 16, an X-crossing in the track assignment between edges both from some $E'_\ell$ (or both from some $E''_\ell$) implies that these edges are nested in $\sigma$. Since no two edges in $E_\ell$ are nested in $\sigma$, the set $\{E'_\ell, E''_\ell : 1 \leq \ell \leq q\}$ defines an edge $2q$-colouring with no monochromatic X-crossing in the track assignment. Thus we have a $(2q, c)$-track layout of $G$. □

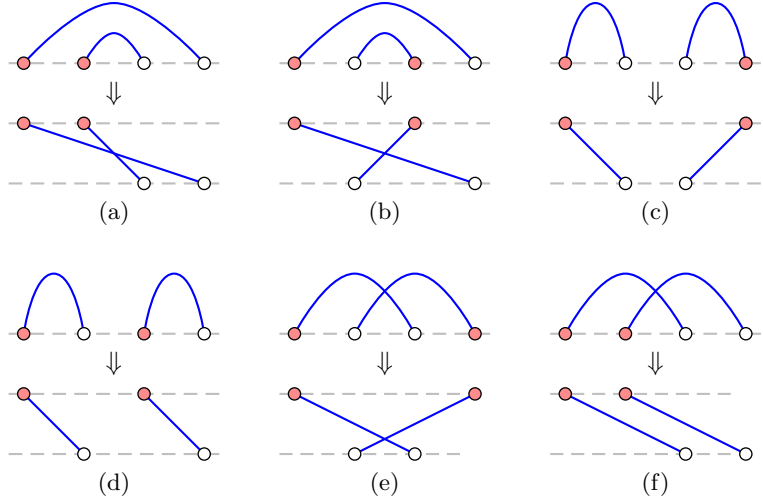

Figure 6: From a linear layout to a track layout: (a)-(b) nested, (c)-(d) disjoint, (e)-(f) crossing.

Lemma 17 is similar to a result by Pemmaraju [49] which says that a queue layout can be 'separated' by a vertex colouring, although the proof by Pemmaraju, which is based on the characterisation of 1-queue graphs as 'arched levelled planar', is much longer. Pemmaraju used separated queue layouts to prove the next result, which follows from Lemmata 13 and 17.

**Theorem 6.** [49] *Stack-number is bounded by queue-number for bipartite graphs. In particular,* $\mathsf{sn}(G) \leq 2\,\mathsf{qn}(G)$ *for every bipartite graph $G$.* □

The *2-track thickness* of a bipartite graph $G$ is the minimum $k$ such that $G$ has a $(k, 2)$-track layout (see [27]). Lemmata 13 and 17 imply:

**Theorem 7.** *Queue-number and 2-track thickness are tied for bipartite graphs. In particular, for every bipartite graph $G$, if $\mathsf{qn}(G) \leq k$ then $G$ has a $(2k, 2)$-track layout, and if $G$ has a $(k, 2)$-track layout then $\mathsf{qn}(G) \leq k$.* □

In our companion paper [30], we prove that every $q$-queue graph is $4q$-colourable. Thus Lemma 17 implies:

**Corollary 4.** *Every $q$-queue graph has a $(2q, 4q)$-track layout.* □

The next corollary of Theorem 2 and Lemma 17 is by Dujmović *et al.* [25].

**Corollary 5.** [25] *Every $q$-queue graph $G$ with acyclic chromatic number $\chi_{\mathrm{a}}(G) \leq c$ has track-number* $\mathsf{tn}(G) \leq c(2q)^{c-1}$. □

Nešetřil and Ossona de Mendez [47] proved that every proper minor-closed graph family has bounded acyclic chromatic number. Thus Corollary 5 implies that for every proper minor-closed graph family, track-number is bounded by queue-number [25]. However, this does not imply that track-number is bounded by queue-number for *all* graphs. For example, in our companion paper [27] we prove that there are 2-queue graphs with unbounded clique minors, for which we cannot apply the result of Nešetřil and Ossona de Mendez [47]. By Theorem 3 and Corollary 4, we have the following result which is qualitatively stronger than Corollary 5.

**Theorem 8.** *Track-number is bounded by queue-number. In particular, every graph $G$ with queue-number $\mathsf{qn}(G) \leq q$ has track-number $\mathsf{tn}(G) \leq 4q \cdot 4^{q(2q-1)(4q-1)}$.* □

Theorem 8 and (1) imply:

**Corollary 6.** *Acyclic chromatic number is bounded by queue-number. In particular, every graph $G$ with queue-number $\mathsf{qn}(G) \leq q$ has acyclic chromatic number $\chi_{\mathrm{a}}(G) \leq 4q \cdot 4^{q(2q-1)(4q-1)}$.* □

Again the converse of Corollary 6 is not true. In particular, $\mathsf{qn}(K_n'') \in \Omega(n^{1/4})$ [27], but $\chi_{\mathrm{a}}(K_n'') = 3$ for $n \geq 3$ (see Section 4). Thus queue-number is not bounded by acyclic chromatic number.

Lemma 10 and Corollary 6 imply:

**Theorem 9.** *Every $q$-queue graph has a $(4q \cdot 4^{q(2q-1)(4q-1)})q^{4q \cdot 4^{q(2q-1)(4q-1)}-1}$-colouring in which each bichromatic subgraph is contained in a single queue, and is thus (arched-levelled) planar.* □

Theorem 8 and Lemma 14 prove the following result, which is one of the main contributions of the paper.

**Theorem 10.** *Queue-number and track-number are tied.* □

In the case of 1-queue graphs, much improved bounds can be obtained. Di Giacomo and Meijer [22] proved that every 1-queue graph has a 5-track layout, and that there exists a 1-queue graph with track-number at least 4. We now prove that the lower bound is the right answer.

**Theorem 11.** *Every 1-queue graph has a $(2,3)$-track layout and a 4-track layout.*

To prove Theorem 11 we will use the following characterisation of 1-queue graphs that may be of independent interest. It is similar but not the same as the characterisation in terms of 'arched levelled' planar graphs due to Heath and Rosenberg [41].

**Lemma 18.** *A graph $G$ has a 1-queue layout if and only if $G$ has a track layout $\{V_i : 1 \leq i \leq t\}$ with maximum span two, such that for every edge $vw \in E(G)$ that has span two $(v \in V_i, w \in V_{i+2})$, $w$ is the first vertex in $V_{i+2}$, and there is no edge $xy \in E(G)$ with $v <_i x \in V_i$ and $y \in V_{i+1}$.*

*Proof.* Suppose that $T = \{V_i : 1 \leq i \leq t\}$ is a track layout of a graph $G$ that satisfies the conditions of the lemma. Let $\sigma$ be the vertex ordering $(V_1, V_2, \ldots, V_t)$. Suppose that there is an edge $xy$ nested inside another edge $vw$ in $\sigma$. Without loss of generality, $v <_\sigma x <_\sigma y <_\sigma w$. By the proof of Lemma 14, edges that have the same span in $T$ are not nested in $\sigma$. Thus the span of $xy$ is one, and the span of $vw$ is two. Hence $v \in V_i$ and $w \in V_{i+2}$ for some $1 \leq i \leq t-2$. By assumption, $w$ is the first vertex in its track. Thus $x \in V_i$ and $y \in V_{i+1}$. But this contradicts the assumption about $T$. Thus no two edges are nested in $\sigma$, and we have a 1-queue layout of $G$.

Now suppose that $\sigma$ is the vertex ordering in a 1-queue layout of a graph $G$. Partition the vertices into non-empty independent sets $V_1, V_2, \ldots, V_t$ such that $\sigma = (V_1, V_2, \ldots, V_t)$, and for all $2 \leq i \leq t$, there exists an edge from the first vertex in $V_i$ to some vertex in $V_{i-1}$. Such a partition can be computed greedily as follows. First let $V_1$ be the largest independent set at the start of $\sigma$. Then for all $i = 2, 3, \ldots, t$, let $V_i$ be the largest independent set starting with the vertex immediately after the last vertex in $V_{i-1}$. We claim that $T = \{(V_i, \sigma) : 1 \leq i \leq t\}$ is the desired track layout.

Since $\sigma$ has no nested edges, by Lemma 16, $T$ is a track layout with no X-crossing. For all $s \geq 3$, there is no edge in any $G[V_i, V_{i+s}]$, as otherwise it would be nested in $\sigma$ with the edge from the first vertex in $V_{i+2}$ to some vertex in $V_{i+1}$. Thus the track layout has span at most two. If an edge $vw$ has span two with $v \in V_i$ and $w \in V_{i+2}$, then $w$ is the first vertex in $V_{i+2}$, as otherwise $vw$ would be nested with the edge between the first vertex in $V_{i+2}$ and some vertex in $V_{i+1}$. Moreover, there is no edge $xy \in E(G)$ with $v <_i x \in V_i$ and $y \in V_{i+1}$, as otherwise $xy$ would be nested inside $vw$ in $\sigma$. $\square$

*Proof of Theorem 11.* Let $T_1 = \{V_i : i \geq 0\}$ be the track layout of $G$ from Lemma 18. Since $T_1$ has maximum span at most two, $G$ has a $(2,3)$-track layout and a 5-track layout by Lemma 7. Let $T_2 = \{W_0, W_1, W_2, W_3\}$ be the track assignment obtained by wrapping $T_1$ modulo four; that is, $W_j = (V_j, V_{j+4}, V_{j+8}, \ldots)$ for $j \in \{0, 1, 2, 3\}$. An edge that has span one in $T_1$ has span one or three in $T_2$. An edge that has span two in $T_1$ has span two in $T_2$. Suppose for the sake of contradiction that two edges $vw$ and $xy$ form an X-crossing in $T_2$. As in Lemma 5, two edges that have span one in $T_1$, do not form an X-crossing in $T_2$. Two edges, one with span one and the other with span two in $T_1$ do not form an X-crossing in $T_2$, as they have different spans in $T_2$. Thus $vw$ and $xy$ both have span two in $T_1$. Thus, by the above observation and without loss of generality, $v$ is the first vertex in some $V_i$ and $x$ is the first vertex in some $V_j$ with $i < j$. Moreover, $j \geq i + 2$ since $v, w, x$ and $y$ belong to two tracks of $T_2$. Thus either $y \in V_\ell$ such that $\ell > i$ or $y \in V_i$ and $y > v$. In either case, as in Lemma 5, $vw$ and $xy$ do not form an X-crossing in $T_2$, which is a contradiction, as desired. $\square$

Theorem 11 and (1) imply:

**Corollary 7.** *Every 1-queue graph $G$ has acyclic chromatic number $\chi_a(G) \leq 4$.* $\square$

Consider a vertex ordering $\sigma = (v_1, v_2, \ldots, v_n)$ of a graph $G$. For each edge $v_i v_j \in E(G)$, let the *width* of $v_i v_j$ in $\sigma$ be $|i - j|$. The *band-width* of $\sigma$ is the maximum width of an edge of $G$ in $\sigma$. The *band-width* of $G$, denoted by $\mathsf{bw}(G)$, is the minimum band-width of a vertex ordering of $G$.

**Lemma 19.** *Every graph $G$ with band-width $\mathsf{bw}(G)$ has track-number $\mathsf{tn}(G) \leq \mathsf{bw}(G) + 1$.*

*Proof.* Let $\sigma = (v_0, v_1, \ldots, v_{n-1})$ be a vertex ordering of $G$ with band-width $b = \mathsf{bw}(G)$. For each $0 \leq \ell \leq b$, let $V_\ell = \{v_i : i \equiv \ell \mod (b+1)\}$. Not only is $\{V_\ell : 0 \leq \ell \leq b\}$ a vertex colouring of $G$, but for every edge $vw$, if there is a vertex $x$ with $v <_\sigma x <_\sigma w$, then all three vertices are in distinct colour classes. Thus, it follows from Lemma 16 that with each $V_\ell$ ordered by $\sigma$, there is no X-crossing. $\square$

Note that Lemma 19 is in fact weaker than the bound due to Dujmovi'c *et al.* [25, 26] that track-number is at most one more than the path-width. However, we consider that this particularly simple proof deserves mention.

# 6 Geometric Thickness

The *geometric thickness* of a graph $G$, denoted by $\overline{\theta}(G)$, is the minimum number of colours such that $G$ can be drawn in the plane with edges as coloured straight-line segments, and monochromatic edges do not cross [23, 31, 42]. Stack-number (when viewed as book-thickness) is equivalent to geometric thickness with the additional requirement that the vertices are in convex position [6]. Thus $\overline{\theta}(G) \leq \mathsf{sn}(G)$ for every graph $G$. While it is an open problem whether stack number is bounded by track-number or by queue-number (see our companion paper [27]), we prove the weaker results that geometric thickness is bounded by track-number, and geometric thickness is bounded by queue-number.

**Theorem 12.** *Geometric thickness is bounded by track-number. In particular, every $(k,t)$-track graph $G$ has geometric thickness $\overline{\theta}(G) \leq k\lceil \frac{t}{2}\rceil \lfloor \frac{t}{2}\rfloor$.*

*Proof.* Let $p \geq t$ be a prime. Position the $j$-th vertex in the $i$-th track at $(i, pj + (i^2 \bmod p))$. Wood [55] proved that in this layout no three vertices are collinear, unless all three are in a single track. Since a track is an independent set, the only vertices that an edge intersects are its own endpoints. The vertices in each track are positioned on a line parallel to the $Y$-axis, in the order defined by the track layout. Thus monochromatic edges between any pair of tracks do not cross. If we let each pair of tracks use a distinct palette of $k$ edge colours, then we obtain a drawing of $G$ with $k\binom{t}{2}$ edge colours, such that monochromatic edges do not cross. That is, $\overline{\theta}(G) \leq k\binom{t}{2}$.

This bound can be improved by partitioning the pairs of tracks into sets that can use the same palette of $k$ colours. This amounts to edge-colouring the complete graph $K_t$ with a fixed vertex ordering $(v_1, v_2, \ldots, v_t)$, so that overlapping edges receive distinct colours. To this end, define a partial order on $E(K_t)$ as follows. For all edges $v_iv_j$ and $v_av_b$ (with $i < j$ and $a < b$), let $v_iv_j \prec v_av_b$ if $j \leq a$. Clearly $\preceq$ is a partial order on $E(K_t)$, such that distinct edges are overlapping if and only if they are incomparable under $\preceq$. By Dilworth's Theorem [24], there is a partition of $E(K_t)$ into $r$ sets, each pairwise non-overlapping, where $r$ is the largest set of pairwise overlapping edges. Clearly $r = \lceil \frac{t}{2}\rceil \lfloor \frac{t}{2}\rfloor$. For each such set, assign a distinct palette of $k$ colours to the edges between pairs of tracks corresponding to edges of $K_t$ in this set. In total we have $kr$ edge colours, and $\overline{\theta}(G) \leq kr = k\lceil \frac{t}{2}\rceil \lfloor \frac{t}{2}\rfloor$. □

Theorem 12 and Lemma 17 imply:

**Corollary 8.** *Every $q$-queue $c$-colourable graph $G$ has geometric thickness $\overline{\theta}(G) \leq 2q\lceil \frac{c}{2}\rceil \lfloor \frac{c}{2}\rfloor$.* □

Theorem 12 and Corollary 4 imply:

**Corollary 9.** *Geometric thickness is bounded by queue-number. In particular, every graph $G$ has geometric thickness $\overline{\theta}(G) \leq 8\,\mathsf{qn}(G)^3$.* □

Note that queue-number is not bounded by geometric thickness. For example, the graph $K'_n$ obtained from $K_n$ by subdividing every edge once has geometric thickness two [32] but has queue-number $\Theta(\sqrt{n})$ [27]. Similarly, acyclic chromatic number is not bounded by geometric thickness, since $\chi_{\mathrm{a}}(K'_n) \in \Theta(\sqrt{n})$ [54]. In fact, in Lemma 20 below we prove a stronger result that provides a counterpoint to Theorems 4 and 9.

Let $G$ be a graph with a straight line drawing in the plane with the vertices in convex position. Suppose that there is a $k$-colouring of the edges so that monochromatic edges do not cross. Then

Theorem 4 implies $G$ has a vertex colouring with $80^{\binom{k}{2}}$ colours so that each bichromatic subgraph is plane (that is, no two edges cross). Theorem 9 gives a similar result for $k$-queue graphs. This type of result does not extend to the case of graphs with geometric thickness $k \geq 2$. Again our example is $K'_n$ which has geometric thickness two [32].

**Lemma 20.** *For every $c \in \mathbb{N}$ there is an $n \in \mathbb{N}$, such that in every vertex $c$-colouring of $K'_n$ there is a bichromatic subgraph that is not planar.*

To prove Lemma 20 we need the following lemma.

**Lemma 21.** *Let $\{V_i : 1 \leq i \leq c\}$ be a vertex vertex $c$-colouring of a graph $G$ such that each bichromatic subgraph $G[V_i, V_j]$ has acyclic chromatic number at most $k$. Then $G$ has acyclic chromatic number $\chi_a(G) \leq c \cdot k^{c-1}$.*

*Proof.* For each vertex $v \in V_i$ and for all $j \neq i$, let $\phi_j(v)$ be the colour assigned to $v$ in an acyclic $k$-colouring of $G[V_i, V_j]$. Colour $v$ by the vector $\phi(v) = (i; \phi_1(v), \ldots, \phi_{i-1}(v), \phi_{i+1}(v), \ldots, \phi_c(v))$. If $C$ is a bichromatic cycle between colour classes $(i; \lambda_1, \ldots, \lambda_{i-1}, \lambda_{i+1}, \ldots, \lambda_c)$ and $(j; \gamma_1, \ldots, \gamma_{j-1}, \gamma_{j+1}, \ldots, \gamma_c)$, then $C$ is a bichromatic cycle between colour classes $\lambda_j$ and $\gamma_i$ in the acyclic colouring of $G[V_i, V_j]$. Thus $\phi$ is an acyclic colouring of $G$. The number of colours is $c \cdot k^{c-1}$. □

*Proof of Lemma 20.* Suppose on the contrary that there is a $c \in \mathbb{N}$ such that every $K'_n$ has a vertex $c$-colouring in which every bichromatic subgraph is planar. By Lemma 21 and since planar graphs have acyclic 5-colourings [10], every $K'_n$ has a $c \cdot 5^{c-1}$-acyclic colouring. However, the acyclic chromatic number of $K'_n$ is $\Omega(\sqrt{n})$ [54]. Thus we obtain a contradiction for sufficiently large $n$. □

## 7 Planar Graphs

Whether planar graphs have bounded track-number is probably the most important open problem in the field. A crossing-free drawing of a graph in the plane in which all the vertices are on the boundary of the outerface is called *outerplanar*. A graph is *outerplanar* if it has an outerplanar drawing. In this section we prove bounds on the track-number of outerplanar graphs, and prove the best known lower bound on the track-number of planar graphs.

**Lemma 22.** *Every outerplanar graph has a 5-track layout.*

*Proof.* We proceed by induction on $n$ with the following hypothesis: Every maximal outerplanar graph $G$ on $n \geq 2$ vertices has a straight-line outerplanar drawing (in which the coordinates of each vertex $v$ are denoted by $(X(v), Y(v))$) such that:

- $Y(v) \in \mathbb{Z}$ for every vertex $v \in V(G)$,

- $|Y(v) - Y(w)| \in \{1, 2\}$ for every edge $vw \in E(G)$,

- the boundary of the drawing is a strictly monotone polygon; that is, every vertical line intersects the boundary in at most two places.

The result will follow since this drawing obviously defines a track layout of $G$ with span two, which can be wrapped onto five tracks by Lemma 7(b). The basis of the induction with $n = 2$ is trivial. Every maximal outerplanar graph on $n \geq 3$ vertices has a vertex $v$ that is adjacent to exactly two vertices $u$ and $w$, such that $uw$ is an edge on the boundary. Let $G' = G \setminus v$. Then $G'$ is also maximal outerplanar. By induction, $G'$ has the desired drawing. The *upper envelope* of the drawing is the portion of the boundary that is visible from $(0, +\infty)$, and the *lower envelope* of the drawing is the portion of the boundary that is visible from $(0, -\infty)$. By the third invariant, every edge that is on the boundary of the drawing is on the upper or lower envelope. Without loss of generality $Y(u) < Y(w)$. As illustrated in Figure 7, position $v$ in the drawing of $G'$ as follows. (For an edge on both envelopes we can use either rule.)

Case (a). $uw$ is on the upper envelope and $Y(w) = Y(u) + 1$: Position $v$ at $(\frac{1}{2}X(w) + \frac{1}{2}X(u), Y(w) + 1)$.
Case (b). $uw$ is on the upper envelope and $Y(w) = Y(u) + 2$: Position $v$ at $(\frac{3}{4}X(u) + \frac{1}{4}X(w), Y(u) + 1)$.
Case (c). $uw$ is on the lower envelope and $Y(w) = Y(u) + 1$: Position $v$ at $(\frac{1}{2}X(u) + \frac{1}{2}X(w), Y(u) - 1)$.
Case (d). $uw$ is on the lower envelope and $Y(w) = Y(u) + 2$: Position $v$ at $(\frac{3}{4}X(w) + \frac{1}{4}X(u), Y(u) + 1)$.

Draw the edges $vu$ and $vw$ straight. It is simple to check that the invariants are maintained. □

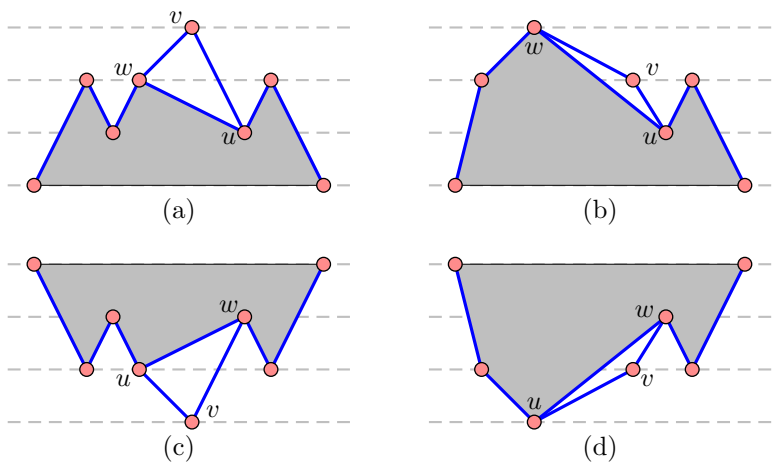

Figure 7: Construction of a track layout of an outerplanar graph in Lemma 22.

We now consider lower bounds on the track-number of outerplanar and planar graphs.

**Lemma 23.** *There is an outerplanar graph $H$ with track-number $\mathsf{tn}(H) \geq 4$.*

*Proof.* Let $H_1, H_2, \ldots, H_5$ be copies of $K_3$. Nominate a vertex $w_j$ of each $H_j$. Construct $H$ by adding an additional vertex $v$ adjacent to each $w_j$. Clearly $H$ is outerplanar. Suppose that $\mathsf{tn}(H) \leq 3$. Without loss of generality, $v$ is in track 1. Thus $\{w_1, w_2, w_3, w_4, w_5\}$ are in tracks 2 and 3. Hence there are three $w_j$ vertices in a single track. Without loss of generality, $w_1 < w_2 < w_3$ in track 2. One vertex $x$ of $H_2$ is in track 1, since $H_2 = K_3$. But this implies that $w_2 x$ forms an X-crossing with $w_1 v$ or $w_3 v$. Hence $\mathsf{tn}(H) \geq 4$. □

Whether every outerplanar graph has a 4-track layout is an interesting open problem. We conjecture that a large enough outerplanar graph whose weak dual is the 'cubic' tree has track-number 5.

**Lemma 24.** *For every outerplanar graph $H$, there is a planar graph $G$ with track-number $\mathsf{tn}(G) \geq \mathsf{tn}(H) + 3$.*

*Proof.* We construct $G$ incrementally. Start with an edge $v_1 v_2$. Let $t = \mathsf{tn}(H)$ and $n = 2t + 1$. Add $n$ new vertices $\{w_1, w_2, \ldots, w_n\}$ each adjacent to both $v_1$ and $v_2$. Let $H_1, H_2, \ldots, H_n$ be copies of $H$. For all $1 \leq j \leq n$, add an edge between $w_j$ and every vertex of $H_j$. As illustrated in Figure 8, $G$ is planar. Suppose that $G$ has a $(t + 2)$-track layout. Without loss of generality, $v_i$ is in track $i$. Thus $\{w_1, w_2, \ldots, w_n\}$ are in tracks $\{3, 4, \ldots, t+2\}$. Hence there are three $w_j$ vertices in a single track. Without loss of generality, $w_1 < w_2 < w_3$ in track 3. No vertex $x$ of $H_2$ is in track 1 or 2, as otherwise $xw_2$ would form an X-crossing with one of $\{v_1 w_1, v_1 w_3, v_2 w_1, v_2 w_3\}$. No vertex $x$ of $H_2$ is in track 3, since $x$ and $w_2$ are adjacent, and $w_2$ is in track 3. Thus all the vertices of $H_2$ are in tracks $\{4, 5, \ldots, t+2\}$, implying $\mathsf{tn}(H) \leq t - 1$, which is a contradiction. Therefore $\mathsf{tn}(G) \geq t + 3$. □

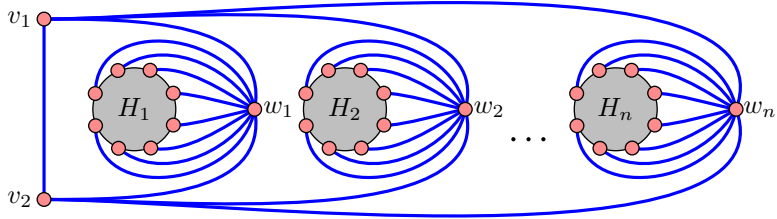

Figure 8: The graph $G$ in Lemma 24.

Lemmata 23 and 24 imply:

**Theorem 13.** *There is a planar graph $G$ with track-number $\mathsf{tn}(G) \geq 7$.* □

The best previous lower bound for the track-number of a planar graph was six due to Giuseppe Liotta and the third author [unpublished].

**Corollary 10.** *There is a planar graph with 'improper track-number' at least $7$.*

*Proof.* Let $G$ be the planar graph from Theorem 13. Let $G'$ be the graph obtained from $G$ by the following construction. For each edge $vw$ of $G$, add six new vertices to $G'$, each adjacent to $v$ and $w$. Clearly $G'$ is planar. Dujmović *et al.* [25] proved that if $G'$ has an improper 6-track layout, then $G$ has a (proper) 6-track layout. Thus $G'$ has no improper 6-track layout by Theorem 13. □

## 8  Computational Complexity

We conclude with some open problems regarding the computational complexity of determining whether a given graph admits a particular type of track layout. Note that there is a simple linear time algorithm to recognise 2-track graphs. Is it $\mathcal{NP}$-complete to recognise $(2,2)$-track graphs? Is it $\mathcal{NP}$-complete to recognise 3-track graphs? Given a vertex ordering $\sigma$ of a graph $G$, is it $\mathcal{NP}$-complete to test if $G$ has a 3-colouring $\{V_1, V_2, V_3\}$ such that $\{(V_1, \sigma), (V_2, \sigma), (V_3, \sigma)\}$ is a track layout?

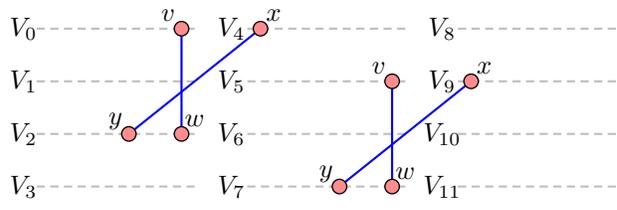